\documentclass{aa} 
\usepackage{natbib}
\usepackage{graphicx}
\usepackage[varg]{txfonts}
\usepackage{epsfig}
\usepackage{times}
\usepackage{rotating}
\usepackage{float}
\usepackage{setspace}
\usepackage{lscape}
\usepackage{tabularx}
\usepackage[usenames]{color}
\usepackage[colorlinks=true,     linkcolor=blue, citecolor=blue, filecolor=blue, urlcolor=blue]{hyperref}
\usepackage{lineno}

\begin{document}

\title{Where do X-ray low surface brightness clusters sit with respect to filaments?}

\authorrunning{S. Zarattini, S. Andreon, and E. Puddu}

\titlerunning{Where do X-ray low surface brightness clusters sit with respect to filaments?}

\author{S. Zarattini\inst{1}, S. Andreon\inst{2}, and E. Puddu\inst{3}}

\institute{Centro de Estudios de F\'isica del Cosmos de Arag\'on (CEFCA), Plaza San Juan 1, 44001 Teruel, Spain 
\and INAF–Osservatorio Astronomico di Brera, Via Brera 28, 20121 Milano, Italy
\and INAF–Osservatorio di Capodimonte, Salita Moiariello 16, 80131 Napoli, Italy \\
email: szarattini@cefca.es
}

\date{\today}
\abstract{}
{The aim of this work is to study the position of gas-rich and gas-poor galaxy clusters within the large-scale structure and, in particular, their distance to filaments.}
{Our sample is built from 29 of the 34 clusters in the X-ray unbiased cluster sample (XUCS), a velocity-dispersion-selected sample for which various properties, including masses, gas fractions, and X-ray surface brightness were available in the literature. We compute the projected distance between each cluster and the spine of the nearest filament with the same redshift and investigate the link between this distance and the previously-mentioned properties of the clusters, in particular with their gas content.}
{The average distance between clusters and filaments is larger for low X-ray surface brightness clusters than for those of high surface brightness, with intermediate brightness clusters being an intermediate case. Also the minimum distance follows a similar trend, with rare cases of low surface brightness clusters found at distances smaller than 2 Mpc from the spine of filaments. However, the Kolmogorov-Smirnov statistical test is not able to exclude the null hypothesis
that the two distributions are coming from the same parent one. 
We speculate that the position of galaxy clusters within the cosmic web could have a direct impact in their gas mass fraction, hence on its X-ray surface brightness, since the presence of a filament can oppose resistance to the outward flow of gas induced by the central AGN and reduce the time required for this gas to fall inward after the AGN is shut. However, a larger sample of clusters is needed in order to derive a statistically-robust conclusion.}
{}


\maketitle

\nolinenumbers

\section{Introduction}
\label{sec:intro}

Studies of the intracluster medium are primarily conducted on galaxy cluster samples selected according to the properties of the intracluster medium itself. This selection can be performed either directly, by detecting the X-ray emission of the gas, or indirectly, by examining its effect on cosmic background photons, a phenomenon known as the Sunyaev–Zeldovich effect \citep[SZ effect,][]{Sunyaev1972}.
It is now accepted that this kind of selections can lead to a biased vision of the entire galaxy cluster population. In fact, for a given mass, X-ray observations can more easily find clusters that are brighter than average, whereas clusters that are fainter than average would be mainly undetected. This bias has often been taken into account \citep[for example in][]{Stanek2006,Pacaud2007,Andreon2011,Andreon2011b,Eckert2011,PlanckIX2011,Andreon2016}, however correcting it is not trivial, since it is based on assumptions on the unseen population. \citet{Andreon2022} showed that the covariance between the position of a cluster in the mass-temperature diagram and its detectability is strong. On the other hand, the luminosity-temperature plane is not affected by the missing population, thus demonstrating that this population behaves differently depending on the analysed scaling relations.

In \citet{Andreon2016}, an unbiased sample was selected using velocity dispersion measurements. Using this sample, called XUCS, the authors demonstrated that the clusters missing from X-ray selection are those with low X-ray surface brightness. Subsequently, \citet{Andreon2017b} showed that differences in surface brightness were related to the gas fraction of clusters. In particular, systems with lower brightness have a smaller gas fraction.

\begin{figure*}
    \centering
    \includegraphics[width=\textwidth]{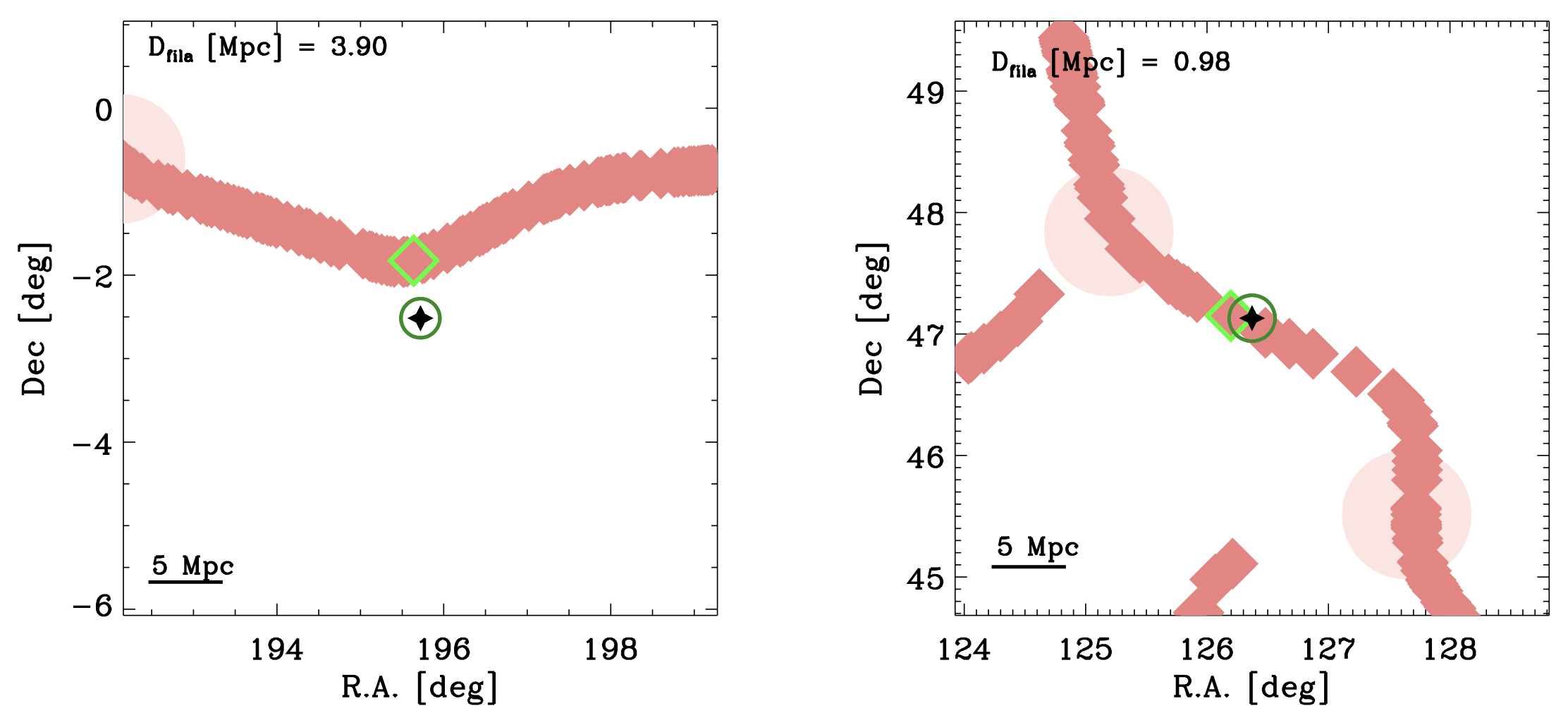}
    \caption{Environment of two clusters, CL1052 (left panel), which is one of the most distant clusters from filaments, and CL3030 (right panel), which is a high surface brightness with typical (median) distance. Our clusters are marked by the black cross in the centre, with the dark green circle representing their virial radii r$_{200}$, whereas  filaments are in coral and intersections (not used in this work) are in rose. The closest filament point, used for measuring the distance, is surrounded by a green rhombus.}   
    \label{fig:example}
\end{figure*}

It was recently found that groups and clusters behave differently depending on their position on the cosmic web. In particular, \citet{Zarattini2023} found that fossil groups (FGs, called in this way for historical reasons, but that include both groups and clusters) are located, on average, farther from both filaments and nodes with respect to non-fossil ones. Moreover, FGs seem unable to survive in the nodes of the cosmic web. Their interpretation is that the position on the cosmic web is the main driver of their evolution, reducing the number of bright galaxies accreted by these systems and affecting the orbits of the infalling galaxies. As a consequence, the merging rate of bright satellites increases, favouring the creation of the large difference in magnitude between the two brightest members of the system. 
Similarly, \citet{Popesso2024} studied the position of GAMA clusters and groups \citep{Driver2022} that were not detected in the early data release of the eROSITA Final Equatorial Depth Survey \citep[eFEDS,][]{Brunner2022}. They found that the eRosita undetected population is more likely to be found in filaments, whereas the population of groups and clusters detected also by eRosita is more likely found in nodes.

It is thus becoming urgent to assess the role of the cosmic web in the evolution of galaxy clusters and groups. In fact, the presence of filaments and nodes could have a direct impact on the gas distribution within the cluster, as it was shown by \citet{Gouin2022} using the IllustrisTNG simulation. In particular, these authors found that the warm and hot intergalactic medium follows the dark matter azimuthal distribution, meaning that the gas is following the filaments in its motion. This effect should leave clear imprints in the mass assembly history, in particular in the formation time, accretion rate, and dynamical state of clusters.
Recently, \citet{Santoni2024} also found in {\tt The Three Hundred} hydrodynamical simulation that cluster masses are related to the connectivity (e.g.  the number of filaments globally connected to a cluster), whereas the dynamical state of clusters seems unrelated to this parameter.

The goal of our work is to test if low surface brightness cluster are found in peculiar positions of the cosmic web and if this position can in some way justify the smaller gas fraction found in these objects. 

Throughout this paper, we assume $\Omega_M=0.3$, $\Omega_\Lambda=0.7$, and $H_0=70$ km s$^{-1}$ Mpc$^{-1}$. 

\section{Sample of clusters and large-scale structure}
\label{sec:sample_lss}

In this Section we introduce the catalogue of filaments and intersections and the X-ray unbiased cluster sample (XUCS) that we use in our analysis.

\subsection{Catalogue of the large-scale structure}
\label{sec:lss}
The catalogue of the large-scale structure that we use in this work was presented in \citet{Chen2016}. It is based on the so called subspace constrained mean shift (SCMS), 
a gradient-based method able to detect filaments through density ridges (smooth curves tracing high-density regions). 

The SCMS method was applied to the Sloan Digital Sky Survey Data Release 12 (SDSS-DR12) spectroscopic data. This is the last data release from the SDSS-III phase and it includes more than one million redshifts from the original SDSS spectrograph, as part of the SDSS-I and SDSS-II, as well as data from the previous year of operations of the SEGUE-2 stellar spectra survey \citep{Alam2015}. The catalogue spans a redshift range between $0.05 < z < 0.7$.

It is worth noting that the SCMS method is tailored to identify filamentary structures and that intersections are defined as where two or more filaments cross with one another. This definition does not precisely identify the nodes of the cosmic web as regions where the density is higher and galaxy clusters reside. \citet{Chen2016} have shown that the distribution of distances from clusters and intersections usually peaks at about two degrees, that is equivalent to about 13 Mpc at z=0.1, the minimum redshift presented in their analysis. For this reason, and to maintain the focus on our main result, we prefer not to include the distance to intersections in our discussion.

\subsection{The X-ray unbiased cluster sample}
\label{sec:sample}

In this work we use the XUCS sample of clusters from \citet{Andreon2016}. This sample is taken from the C4 catalogue of \citet{Miller2005}, that was built by looking for over densities in the seven-dimensional space of position, redshift, and colours obtained from the second data release of the Sloan Digital Sky Survey \citep[][]{Abazajian2004}. The XUCS sample is built by all the C4 clusters with more than 50 spectroscopic members within 1 Mpc, a velocity dispersion $\sigma_v > 500$ km s$^{-1}$, and an additional mass-dependent redshift range ($ 0.05 < z < 0.135$), in order to optimise the X-ray follow up with the Swift satellite. 

The XUCS cluster sample is thus composed by 34 clusters \citep{Andreon2016}, they all have masses estimated using the caustic method \citep{Diaferio1997, Diaferio1999, Serra2011}, which is unaffected by the cluster dynamical status, using more than 100 galaxy velocities per cluster, on average. The masses of the full XUCS sample can be found in \citet{Andreon2016} and the mass range is $13.5 < {\rm log} M_{500}/M_\odot \le 14.6$, with an interquartile range of  $13.9 < {\rm log} M_{500}/M_\odot \le 14.3$ \citep{Andreon2024}. These masses are consistent with hydrostatic masses \citep{Andreon2017} and have an average error of 0.14 dex.

\begin{table}[]
    \caption{Distance to the closest filament, in Mpc, for all the clusters used in this work. The Id is the same as in Table 1 of \citet{Andreon2016}}
    \centering
    \begin{tabular}{lc|lc}
    \hline
    Id & Distance & Id & Distance \\
     & [Mpc]  &   & [Mpc] \\
    \hline
    CL1001 &  0.98 &  CL1067 & 2.77  \\
    CL1009 &  1.56 &  CL1073 & 0.24 \\
    CL1011 &  6.14 &  CL1120 &  4.63 \\
    CL1014 &  1.03 &  CL1132 &  0.78\\
    CL1015 &  2.72 &  CL1209 &  2.34 \\
    CL1018 &  0.40 &  CL3000 &  0.98 \\
    CL1020 &  2.22 &  CL3009 &  3.89 \\
    CL1022 &  1.50 &  CL3013 &  0.49 \\
    CL1030 &  2.06 &  CL3020 &  1.39 \\
    CL1033 &  0.67 &  CL3023 &  0.99 \\
    CL1038 &  3.15 &  CL3030 &  0.98 \\
    CL1039 &  0.98 &  CL3046 &  2.06 \\
    CL1041 &  0.82 &  CL3049 &  3.16 \\
    CL1047 &  0.38 &  CL3053 &  1.73 \\
    CL1052 &  3.90 &               &          \\
    
    \end{tabular}
    \label{tab:distances}
\end{table}

In \citet{Puddu2022} the XUCS clusters were classified in three categories based on the X-ray surface brightness within $r_{500}$, SB$_{\rm{X}}$. Clusters are classified of high surface brightness if $\rm{log \, SB_{X}} \ge 43.35$ erg s$^{-1}$ Mpc$^{-2}$ and they have low surface brightness if $\rm{log \, SB_{X}} \le 42.55$ erg s$^{-1}$ Mpc$^{-2}$. 

Those in the intermediate range are considered as intermediate cases. Clusters with low surface brightness for their mass are found to have also low richness  for their mass \citep{Puddu2022} and,
according to simulations, experienced a strong AGN activity at early ages \citep{Ragagnin2022}.

For this work we only use those XUCS clusters with $120 < {\rm R.A.} < 240$ degrees and $-5 < {\rm Dec} < 70$ degrees because the \citet{Chen2016} catalogue is uniform in this region only. 
As a consequence, the final XUCS sample that accomplishes this further constraint is reduced to 29 clusters.

\section{Results}
\label{sec:results}
The method adopted in this work for measuring the distances between clusters and filaments is similar to the one presented in \citet{Zarattini2022,Zarattini2023}. The filament catalogue is divided into thin slices of redshift, starting from $z=0.05$ and with a thickness of $0.005$ \citep[equivalent to $\sim 1500$ km s$^{-1}$,][]{Chen2016}. As a first step, we identify the slice corresponding to the redshift of the cluster.
The filaments are described with a series of points with R.A. and Dec. These points are not contiguous, but they are usually dense enough to identify the spine of each filament (see Fig. \ref{fig:example}). Since the slices are very thin, we assume that the cluster and the filaments are exactly at the same
redshift and we define the distance between them as the minimum projected mutual distance. This distance is reported in Table \ref{tab:distances}.

\begin{figure}[t]
    \centering
    \includegraphics[width=0.45\textwidth,trim=60 0 40 20]{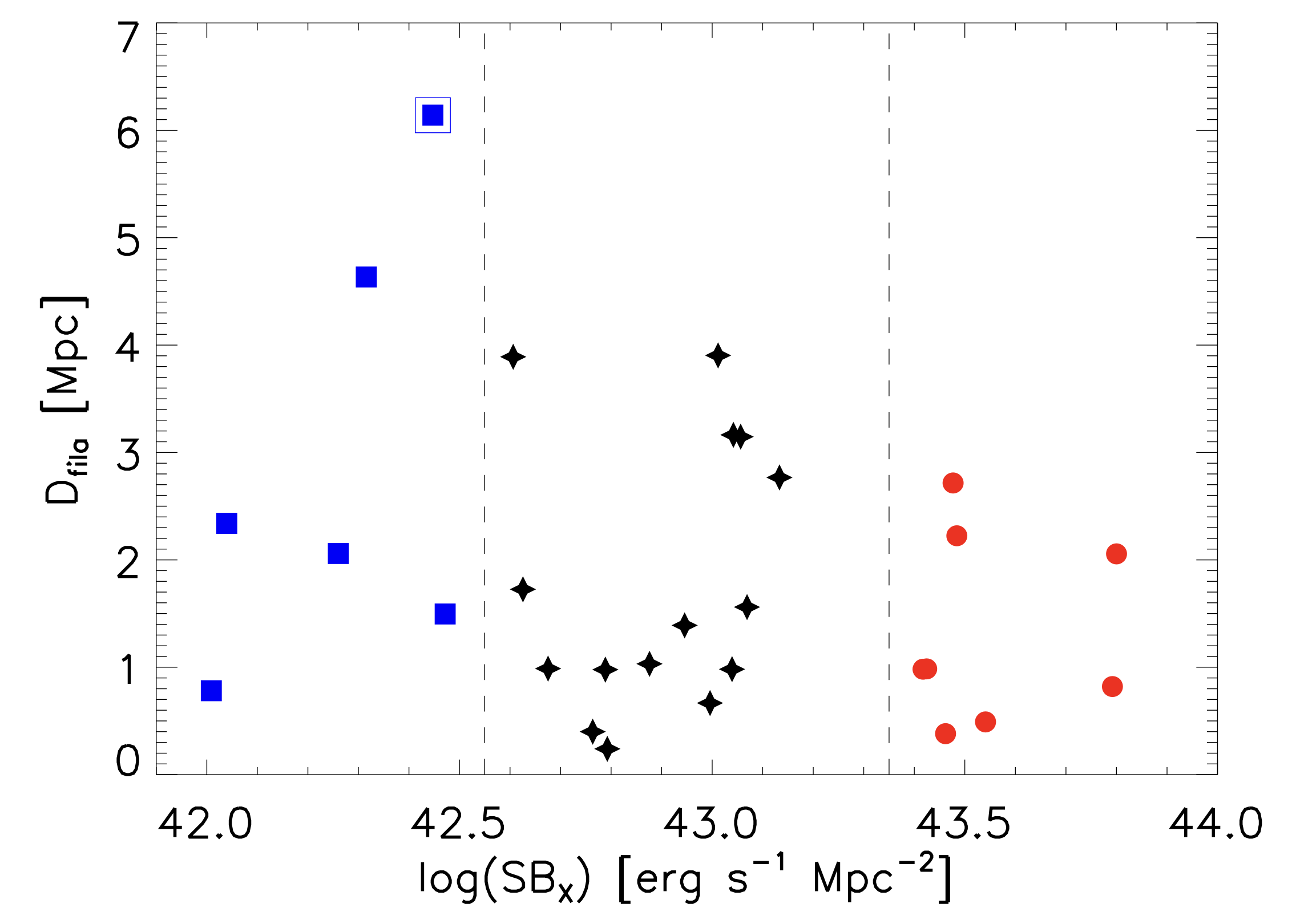}
    \caption{Distance between our clusters and the nearest filament as a function of their X-ray surface brightness SB$_{\rm X}$. 
    The figure is colour-coded according to the X-ray surface brightness: high-surface brightness clusters are shown with red circles, low-surface brightness with blue squares, and intermediate cases with black stars. The farthest object in our sample is highlighted with a blue open square and it refers to CL1011, a peculiar case that we analyse in details in Appendix A}
    \label{fig:SBx}
\end{figure}

In Fig. \ref{fig:SBx} we show the distance between our clusters and the closest filament, expressed in Mpc, as a function of the X-ray surface brightness. It can be seen that the plot has a sort of triangular shape, with high surface brightness clusters found closer to filaments than the low surface brightness ones.
In fact, no gas-rich cluster is found at more than 3 Mpc from filaments. On the other hand, gas-intermediate and gas-poor clusters (black and blue points, respectively) can be found at larger distances from filaments. 

Gas-rich and gas-intermediate clusters often have distances smaller than half Mpc, whereas gas-poor clusters are rarely found at distance smaller than 2 Mpc.

There is also a sort of a dichotomy in the cluster population of Fig. \ref{fig:SBx}. In fact, there seems to be a diagonal gap covering all the surface brightness range. We modelled this possible dichotomy in Appendix~\ref{sec:dicotomy}, however we were unable to find any impact of this possible dichotomy on cluster global properties (halo mass, magnitude of the BCGs, magnitude gap, and X-ray temperature).

\begin{figure}
    \centering
    \includegraphics[width=0.42\textwidth, trim= 70 0 60 10]{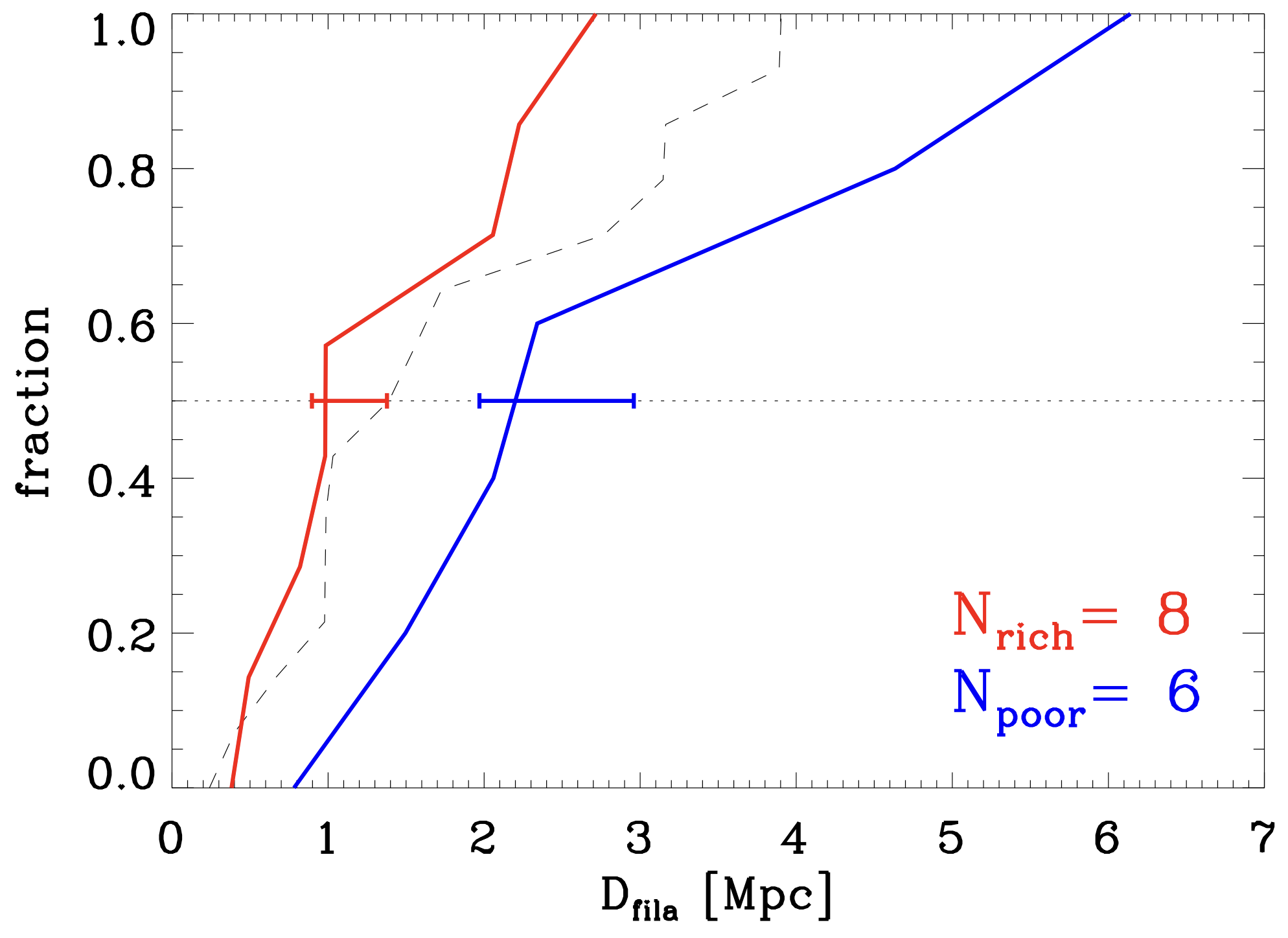}
    \caption{Cumulative distribution of the distance to the nearest filament for clusters of large (red solid line),  intermediate (black dashed line), and low surface brightness (blue solid line). The horizontal dotted line indicates the median of each distribution and the limits of the error bars are the 25 and 75 percentile of each distribution. }
    \label{fig:cumulative}
\end{figure}

Finally, we plot in Fig. \ref{fig:cumulative} the cumulative distribution of  the distance from filaments for the three classes of objects. The two subsamples of gas-rich and gas-poor systems seem to follow different distributions. In particular, their median values are ${\rm D_{fila,rich}} =0.98$ Mpc and ${\rm D_{fila,rich}} =2.20$ Mpc. In Fig. \ref{fig:cumulative} we also reported the 25 and 75 percentiles uncertainties.
However, the Kolmogorov-Smirnov (KS), with a p-value of 0.35, is unable to reject the null hypothesis that the two samples are drawn from the same parent distribution. Using the model introduced in Appendix~\ref{sec:dicotomy}, we found that the dependence of the distance from the filament on the cluster brightness is only about $1.5\sigma$ significant. Therefore, our evidence remain tentative only and requires a larger sample to be established with larger significance.

\section{Discussion and conclusions}
\label{sec:discussion}

In this work we compute the distance between a sample of clusters and the filaments of the cosmic web. We divide our sample in three subsamples according to their X-ray surface brightness and we mainly focus our attention to the subsamples of high- and low-surface brightness clusters, with the third subsample being an intermediate case.
As mentioned, X-ray surface brightness is tightly related to gas fraction \citep{Andreon2017,Ragagnin2022}.

We found that there seems to be a trend for which, on average, gas-poor systems are find at larger distances than gas-rich ones. However the difference is statistically inconclusive, probably due to the small size of the sample, since the gas-rich clusters are 8 and gas-poor ones are 6.

The position of clusters within the cosmic web could have an impact on the mechanisms affecting their gas reservoir. 
The low gas fraction of some of our clusters \citep{Andreon2017b} and the low values measured by
\citet{Hadzhiyska2024} and \citet{Bigwood2024}, all indicate that a considerable fraction (up to 50\%) of the gas mass can be displaced away from the central regions. Indeed, \citet{Bigwood2024} observations constraint the lack to even larger radii and \citet{Hadzhiyska2024} observations show that the gas is much more extended than the dark matter.  Both studies interpret their results as being influenced by AGN feedback, which, according to \citet{Hadzhiyska2024}, is stronger than that observed in the Illustris-TNG cosmological simulation \citep{Nelson2019}, and, as noted by \citet{Bigwood2024}, stronger than in most modern simulations. The scatter in the
gas fraction and its possible dependence from the cluster position in the cosmic web let us speculate that the presence of a filament may in part control the gas flow: the gas driven at large radii by the AGN could find a stronger resistance by the filament, thus reducing the speed and the distance that can be reached by the gas displaced by the AGN feedback. Furthermore, since the gas is closer to the cluster centre than in absence of a filament, the come back would be faster once the AGN feedback is completed. Both effects would make clusters near the spine of filaments more gas rich and clusters further away gas poor, as our observations seem to suggest.

The identification of the mechanisms able to modulate the gas flow from/to large distances, as the ones proposed above, is crucial for understanding the variety of gas fractions in the both the central and non-central regions of galaxy clusters.

Concluding, in this work we study a sample of 29 velocity-dispersion selected clusters classified in X-ray surface brightness classes, which are also gas-fraction classes, with the goal of analysing the large scale structure around these objects and, in particular, their distance from the nearest filament. We find hints that gas-poor clusters are found, on average, at larger distances from filaments than gas-rich ones, although the small sample size
prevented to draw final conclusions. A larger sample will certainly reduce the statistical uncertainty that is affecting our determination.

\begin{acknowledgements}
SZ acknowledges the financial support provided by the Governments of Spain and Arag\'on through their general budgets and the Fondo de Inversiones de Teruel, the Aragonese Government through the Research Group E16\_23R, the Spanish Ministry of Science and Innovation (MCIN/AEI/10.13039/501100011033) and ERDF, A way of making Europe with grant PID2021-124918NB-C44, and from the Spanish Ministry of Science and Innovation and the European Union - NextGenerationEU through the Recovery and Resilience Facility project ICTS-MRR-2021-03-CEFCA. SA and EP acknowledge PRIN-MIUR grant
20228B938N ``Mass and selection biases of galaxy clusters: a multi-probe approach". SA acknowledges INAF grant  ``Characterizing the newly  discovered clusters of low surface  brightness".
\end{acknowledgements}

\bibliography{bibliografia}

\begin{thebibliography}{33}
\expandafter\ifx\csname natexlab\endcsname\relax\def\natexlab#1{#1}\fi

\bibitem[{{Abazajian} {et~al.}(2004){Abazajian}, {Adelman-McCarthy},
  {Ag{\"u}eros}, {Allam}, {Anderson}, {Anderson}, {Annis}, {Bahcall}, {Baldry},
  {Bastian}, {Berlind}, {Bernardi}, {Blanton}, {Bochanski}, {Boroski},
  {Briggs}, {Brinkmann}, {Brunner}, {Budav{\'a}ri}, {Carey}, {Carliles},
  {Castander}, {Connolly}, {Csabai}, {Doi}, {Dong}, {Eisenstein}, {Evans},
  {Fan}, {Finkbeiner}, {Friedman}, {Frieman}, {Fukugita}, {Gal}, {Gillespie},
  {Glazebrook}, {Gray}, {Grebel}, {Gunn}, {Gurbani}, {Hall}, {Hamabe},
  {Harris}, {Harris}, {Harvanek}, {Heckman}, {Hendry}, {Hennessy}, {Hindsley},
  {Hogan}, {Hogg}, {Holmgren}, {Ichikawa}, {Ichikawa}, {Ivezi{\'c}}, {Jester},
  {Johnston}, {Jorgensen}, {Kent}, {Kleinman}, {Knapp}, {Kniazev}, {Kron},
  {Krzesinski}, {Kunszt}, {Kuropatkin}, {Lamb}, {Lampeitl}, {Lee}, {Leger},
  {Li}, {Lin}, {Loh}, {Long}, {Loveday}, {Lupton}, {Malik}, {Margon},
  {Matsubara}, {McGehee}, {McKay}, {Meiksin}, {Munn}, {Nakajima}, {Nash},
  {Neilsen}, {Newberg}, {Newman}, {Nichol}, {Nicinski}, {Nieto-Santisteban},
  {Nitta}, {Okamura}, {O'Mullane}, {Ostriker}, {Owen}, {Padmanabhan},
  {Peoples}, {Pier}, {Pope}, {Quinn}, {Richards}, {Richmond}, {Rix}, {Rockosi},
  {Schlegel}, {Schneider}, {Scranton}, {Sekiguchi}, {Seljak}, {Sergey},
  {Sesar}, {Sheldon}, {Shimasaku}, {Siegmund}, {Silvestri}, {Smith},
  {Smol{\v{c}}i{\'c}}, {Snedden}, {Stebbins}, {Stoughton}, {Strauss},
  {SubbaRao}, {Szalay}, {Szapudi}, {Szkody}, {Szokoly}, {Tegmark}, {Teodoro},
  {Thakar}, {Tremonti}, {Tucker}, {Uomoto}, {Vanden Berk}, {Vandenberg},
  {Vogeley}, {Voges}, {Vogt}, {Walkowicz}, {Wang}, {Weinberg}, {West}, {White},
  {Wilhite}, {Xu}, {Yanny}, {Yasuda}, {Yip}, {Yocum}, {York}, {Zehavi},
  {Zibetti}, \& {Zucker}}]{Abazajian2004}
{Abazajian}, K., {Adelman-McCarthy}, J.~K., {Ag{\"u}eros}, M.~A., {et~al.}
  2004, \aj, 128, 502

\bibitem[{{Alam} {et~al.}(2015){Alam}, {Albareti}, {Allende Prieto}, {Anders},
  {Anderson}, {Anderton}, {Andrews}, {Armengaud}, {Aubourg}, {Bailey}, {Basu},
  {Bautista}, {Beaton}, {Beers}, {Bender}, {Berlind}, {Beutler}, {Bhardwaj},
  {Bird}, {Bizyaev}, {Blake}, {Blanton}, {Blomqvist}, {Bochanski}, {Bolton},
  {Bovy}, {Shelden Bradley}, {Brandt}, {Brauer}, {Brinkmann}, {Brown},
  {Brownstein}, {Burden}, {Burtin}, {Busca}, {Cai}, {Capozzi}, {Carnero
  Rosell}, {Carr}, {Carrera}, {Chambers}, {Chaplin}, {Chen}, {Chiappini},
  {Chojnowski}, {Chuang}, {Clerc}, {Comparat}, {Covey}, {Croft}, {Cuesta},
  {Cunha}, {da Costa}, {Da Rio}, {Davenport}, {Dawson}, {De Lee}, {Delubac},
  {Deshpande}, {Dhital}, {Dutra-Ferreira}, {Dwelly}, {Ealet}, {Ebelke},
  {Edmondson}, {Eisenstein}, {Ellsworth}, {Elsworth}, {Epstein}, {Eracleous},
  {Escoffier}, {Esposito}, {Evans}, {Fan}, {Fern{\'a}ndez-Alvar}, {Feuillet},
  {Filiz Ak}, {Finley}, {Finoguenov}, {Flaherty}, {Fleming}, {Font-Ribera},
  {Foster}, {Frinchaboy}, {Galbraith-Frew}, {Garc{\'\i}a},
  {Garc{\'\i}a-Hern{\'a}ndez}, {Garc{\'\i}a P{\'e}rez}, {Gaulme}, {Ge},
  {G{\'e}nova-Santos}, {Georgakakis}, {Ghezzi}, {Gillespie}, {Girardi},
  {Goddard}, {Gontcho}, {Gonz{\'a}lez Hern{\'a}ndez}, {Grebel}, {Green},
  {Grieb}, {Grieves}, {Gunn}, {Guo}, {Harding}, {Hasselquist}, {Hawley},
  {Hayden}, {Hearty}, {Hekker}, {Ho}, {Hogg}, {Holley-Bockelmann}, {Holtzman},
  {Honscheid}, {Huber}, {Huehnerhoff}, {Ivans}, {Jiang}, {Johnson},
  {Kinemuchi}, {Kirkby}, {Kitaura}, {Klaene}, {Knapp}, {Kneib}, {Koenig},
  {Lam}, {Lan}, {Lang}, {Laurent}, {Le Goff}, {Leauthaud}, {Lee}, {Lee},
  {Licquia}, {Liu}, {Long}, {L{\'o}pez-Corredoira}, {Lorenzo-Oliveira},
  {Lucatello}, {Lundgren}, {Lupton}, {Mack}, {Mahadevan}, {Maia}, {Majewski},
  {Malanushenko}, {Malanushenko}, {Manchado}, {Manera}, {Mao}, {Maraston},
  {Marchwinski}, {Margala}, {Martell}, {Martig}, {Masters}, {Mathur},
  {McBride}, {McGehee}, {McGreer}, {McMahon}, {M{\'e}nard}, {Menzel},
  {Merloni}, {M{\'e}sz{\'a}ros}, {Miller}, {Miralda-Escud{\'e}}, {Miyatake},
  {Montero-Dorta}, {More}, {Morganson}, {Morice-Atkinson}, {Morrison},
  {Mosser}, {Muna}, {Myers}, {Nandra}, {Newman}, {Neyrinck}, {Nguyen},
  {Nichol}, {Nidever}, {Noterdaeme}, {Nuza}, {O'Connell}, {O'Connell},
  {O'Connell}, {Ogando}, {Olmstead}, {Oravetz}, {Oravetz}, {Osumi}, {Owen},
  {Padgett}, {Padmanabhan}, {Paegert}, {Palanque-Delabrouille}, {Pan},
  {Parejko}, {P{\^a}ris}, {Park}, {Pattarakijwanich}, {Pellejero-Ibanez},
  {Pepper}, {Percival}, {P{\'e}rez-Fournon}, {P{\'e}rez-R{\`a}fols},
  {Petitjean}, {Pieri}, {Pinsonneault}, {Porto de Mello}, {Prada}, {Prakash},
  {Price-Whelan}, {Protopapas}, {Raddick}, {Rahman}, {Reid}, {Rich}, {Rix},
  {Robin}, {Rockosi}, {Rodrigues}, {Rodr{\'\i}guez-Torres}, {Roe}, {Ross},
  {Ross}, {Rossi}, {Ruan}, {Rubi{\~n}o-Mart{\'\i}n}, {Rykoff},
  {Salazar-Albornoz}, {Salvato}, {Samushia}, {S{\'a}nchez}, {Santiago},
  {Sayres}, {Schiavon}, {Schlegel}, {Schmidt}, {Schneider}, {Schultheis},
  {Schwope}, {Sc{\'o}ccola}, {Scott}, {Sellgren}, {Seo}, {Serenelli}, {Shane},
  {Shen}, {Shetrone}, {Shu}, {Silva Aguirre}, {Sivarani}, {Skrutskie},
  {Slosar}, {Smith}, {Sobreira}, {Souto}, {Stassun}, {Steinmetz}, {Stello},
  {Strauss}, {Streblyanska}, {Suzuki}, {Swanson}, {Tan}, {Tayar}, {Terrien},
  {Thakar}, {Thomas}, {Thomas}, {Thompson}, {Tinker}, {Tojeiro}, {Troup},
  {Vargas-Maga{\~n}a}, {Vazquez}, {Verde}, {Viel}, {Vogt}, {Wake}, {Wang},
  {Weaver}, {Weinberg}, {Weiner}, {White}, {Wilson}, {Wisniewski},
  {Wood-Vasey}, {Ye`che}, {York}, {Zakamska}, {Zamora}, {Zasowski}, {Zehavi},
  {Zhao}, {Zheng}, {Zhou}, {Zhou}, {Zou}, \& {Zhu}}]{Alam2015}
{Alam}, S., {Albareti}, F.~D., {Allende Prieto}, C., {et~al.} 2015, \apjs, 219,
  12

\bibitem[{{Andreon} \& {Hurn}(2010)}]{Andreon2010}
{Andreon}, S. \& {Hurn}, M.~A. 2010, \mnras, 404, 1922

\bibitem[{{Andreon} \& {Moretti}(2011)}]{Andreon2011b}
{Andreon}, S. \& {Moretti}, A. 2011, \aap, 536, A37

\bibitem[{{Andreon} {et~al.}(2016){Andreon}, {Serra}, {Moretti}, \&
  {Trinchieri}}]{Andreon2016}
{Andreon}, S., {Serra}, A.~L., {Moretti}, A., \& {Trinchieri}, G. 2016, \aap,
  585, A147

\bibitem[{{Andreon} {et~al.}(2022){Andreon}, {Trinchieri}, \&
  {Moretti}}]{Andreon2022}
{Andreon}, S., {Trinchieri}, G., \& {Moretti}, A. 2022, \mnras, 511, 4991

\bibitem[{{Andreon} {et~al.}(2024){Andreon}, {Trinchieri}, \&
  {Moretti}}]{Andreon2024}
{Andreon}, S., {Trinchieri}, G., \& {Moretti}, A. 2024, \aap, 686, A284

\bibitem[{{Andreon} {et~al.}(2017{\natexlab{a}}){Andreon}, {Trinchieri},
  {Moretti}, \& {Wang}}]{Andreon2017}
{Andreon}, S., {Trinchieri}, G., {Moretti}, A., \& {Wang}, J.
  2017{\natexlab{a}}, \aap, 606, A25

\bibitem[{{Andreon} {et~al.}(2011){Andreon}, {Trinchieri}, \&
  {Pizzolato}}]{Andreon2011}
{Andreon}, S., {Trinchieri}, G., \& {Pizzolato}, F. 2011, \mnras, 412, 2391

\bibitem[{{Andreon} {et~al.}(2017{\natexlab{b}}){Andreon}, {Wang},
  {Trinchieri}, {Moretti}, \& {Serra}}]{Andreon2017b}
{Andreon}, S., {Wang}, J., {Trinchieri}, G., {Moretti}, A., \& {Serra}, A.~L.
  2017{\natexlab{b}}, \aap, 606, A24

\bibitem[{{Bigwood} {et~al.}(2024){Bigwood}, {Amon}, {Schneider}, {Salcido},
  {McCarthy}, {Preston}, {Sanchez}, {Sijacki}, {Schaan}, {Ferraro},
  {Battaglia}, {Chen}, {Dodelson}, {Roodman}, {Pieres}, {Fert{\'e}}, {Alarcon},
  {Drlica-Wagner}, {Choi}, {Navarro-Alsina}, {Campos}, {Ross}, {Carnero
  Rosell}, {Yin}, {Yanny}, {S{\'a}nchez}, {Chang}, {Davis}, {Doux}, {Gruen},
  {Rykoff}, {Huff}, {Sheldon}, {Tarsitano}, {Andrade-Oliveira}, {Bernstein},
  {Giannini}, {Diehl}, {Huang}, {Harrison}, {Sevilla-Noarbe}, {Tutusaus},
  {Elvin-Poole}, {McCullough}, {Zuntz}, {Blazek}, {DeRose}, {Cordero}, {Prat},
  {Myles}, {Eckert}, {Bechtol}, {Herner}, {Secco}, {Gatti}, {Raveri}, {Kind},
  {Becker}, {Troxel}, {Jarvis}, {MacCrann}, {Friedrich}, {Alves}, {Leget},
  {Chen}, {Rollins}, {Wechsler}, {Gruendl}, {Cawthon}, {Allam}, {Bridle},
  {Pandey}, {Everett}, {Shin}, {Hartley}, {Fang}, {Zhang}, {Aguena}, {Annis},
  {Bacon}, {Bertin}, {Bocquet}, {Brooks}, {Carretero}, {Castander}, {da Costa},
  {Pereira}, {De Vicente}, {Desai}, {Doel}, {Ferrero}, {Flaugher}, {Frieman},
  {Garc{\'\i}a-Bellido}, {Gaztanaga}, {Gutierrez}, {Hinton}, {Hollowood},
  {Honscheid}, {Huterer}, {James}, {Kuehn}, {Lahav}, {Lee}, {Marshall},
  {Mena-Fern{\'a}ndez}, {Miquel}, {Muir}, {Paterno}, {Plazas Malag{\'o}n},
  {Porredon}, {Romer}, {Samuroff}, {Sanchez}, {Sanchez Cid}, {Smith},
  {Soares-Santos}, {Suchyta}, {Swanson}, {Tarle}, {To}, {Weaverdyck}, {Weller},
  {Wiseman}, \& {Yamamoto}}]{Bigwood2024}
{Bigwood}, L., {Amon}, A., {Schneider}, A., {et~al.} 2024, \mnras, 534, 655

\bibitem[{{Brunner} {et~al.}(2022){Brunner}, {Liu}, {Lamer}, {Georgakakis},
  {Merloni}, {Brusa}, {Bulbul}, {Dennerl}, {Friedrich}, {Liu}, {Maitra},
  {Nandra}, {Ramos-Ceja}, {Sanders}, {Stewart}, {Boller}, {Buchner}, {Clerc},
  {Comparat}, {Dwelly}, {Eckert}, {Finoguenov}, {Freyberg}, {Ghirardini},
  {Gueguen}, {Haberl}, {Kreykenbohm}, {Krumpe}, {Osterhage}, {Pacaud},
  {Predehl}, {Reiprich}, {Robrade}, {Salvato}, {Santangelo}, {Schrabback},
  {Schwope}, \& {Wilms}}]{Brunner2022}
{Brunner}, H., {Liu}, T., {Lamer}, G., {et~al.} 2022, \aap, 661, A1

\bibitem[{{Chen} {et~al.}(2016){Chen}, {Ho}, {Brinkmann}, {Freeman},
  {Genovese}, {Schneider}, \& {Wasserman}}]{Chen2016}
{Chen}, Y.-C., {Ho}, S., {Brinkmann}, J., {et~al.} 2016, \mnras, 461, 3896

\bibitem[{{Diaferio}(1999)}]{Diaferio1999}
{Diaferio}, A. 1999, \mnras, 309, 610

\bibitem[{{Diaferio} \& {Geller}(1997)}]{Diaferio1997}
{Diaferio}, A. \& {Geller}, M.~J. 1997, \apj, 481, 633

\bibitem[{{Driver} {et~al.}(2022){Driver}, {Bellstedt}, {Robotham}, {Baldry},
  {Davies}, {Liske}, {Obreschkow}, {Taylor}, {Wright}, {Alpaslan}, {Bamford},
  {Bauer}, {Bland-Hawthorn}, {Bilicki}, {Bravo}, {Brough}, {Casura}, {Cluver},
  {Colless}, {Conselice}, {Croom}, {de Jong}, {D'Eugenio}, {De Propris},
  {Dogruel}, {Drinkwater}, {Dvornik}, {Farrow}, {Frenk}, {Giblin}, {Graham},
  {Grootes}, {Gunawardhana}, {Hashemizadeh}, {H{\"a}u{\ss}ler}, {Heymans},
  {Hildebrandt}, {Holwerda}, {Hopkins}, {Jarrett}, {Heath Jones}, {Kelvin},
  {Koushan}, {Kuijken}, {Lara-L{\'o}pez}, {Lange}, {L{\'o}pez-S{\'a}nchez},
  {Loveday}, {Mahajan}, {Meyer}, {Moffett}, {Napolitano}, {Norberg}, {Owers},
  {Radovich}, {Raouf}, {Peacock}, {Phillipps}, {Pimbblet}, {Popescu}, {Said},
  {Sansom}, {Seibert}, {Sutherland}, {Thorne}, {Tuffs}, {Turner}, {van der
  Wel}, {van Kampen}, \& {Wilkins}}]{Driver2022}
{Driver}, S.~P., {Bellstedt}, S., {Robotham}, A. S.~G., {et~al.} 2022, \mnras,
  513, 439

\bibitem[{{Eckert} {et~al.}(2011){Eckert}, {Molendi}, \&
  {Paltani}}]{Eckert2011}
{Eckert}, D., {Molendi}, S., \& {Paltani}, S. 2011, \aap, 526, A79

\bibitem[{{Gouin} {et~al.}(2022){Gouin}, {Gallo}, \& {Aghanim}}]{Gouin2022}
{Gouin}, C., {Gallo}, S., \& {Aghanim}, N. 2022, \aap, 664, A198

\bibitem[{{Hadzhiyska} {et~al.}(2024){Hadzhiyska}, {Ferraro}, {Ried Guachalla},
  {Schaan}, {Aguilar}, {Battaglia}, {Bond}, {Brooks}, {Calabrese}, {Choi},
  {Claybaugh}, {Coulton}, {Dawson}, {Devlin}, {Dey}, {Doel}, {Duivenvoorden},
  {Dunkley}, {Farren}, {Font-Ribera}, {Forero-Romero}, {Gallardo},
  {Gazta{\~n}aga}, {Gontcho Gontcho}, {Gralla}, {Le Guillou}, {Gutierrez},
  {Guy}, {Hill}, {Hlo{\v{z}}ek}, {Honscheid}, {Juneau}, {Kisner}, {Kremin},
  {Landriau}, {Liu}, {Louis}, {MacCrann}, {de Macorra}, {Madhavacheril},
  {Manera}, {Meisner}, {Miquel}, {Moodley}, {Moustakas}, {Mroczkowski},
  {Naess}, {Newman}, {Niemack}, {Niz}, {Page}, {Palanque-Delabrouille},
  {Partridge}, {Percival}, {Prada}, {Qu}, {Rossi}, {Sanchez}, {Schlegel},
  {Schubnell}, {Sehgal}, {Seo}, {Sif{\'o}n}, {Spergel}, {Sprayberry}, {Staggs},
  {Tarl{\'e}}, {Vargas}, {Vavagiakis}, {Weaver}, {Wollack}, {Zhou}, \&
  {Zou}}]{Hadzhiyska2024}
{Hadzhiyska}, B., {Ferraro}, S., {Ried Guachalla}, B., {et~al.} 2024, arXiv
  e-prints, arXiv:2407.07152

\bibitem[{{Miller} {et~al.}(2005){Miller}, {Nichol}, {Reichart}, {Wechsler},
  {Evrard}, {Annis}, {McKay}, {Bahcall}, {Bernardi}, {Boehringer}, {Connolly},
  {Goto}, {Kniazev}, {Lamb}, {Postman}, {Schneider}, {Sheth}, \&
  {Voges}}]{Miller2005}
{Miller}, C.~J., {Nichol}, R.~C., {Reichart}, D., {et~al.} 2005, \aj, 130, 968

\bibitem[{{Nelson} {et~al.}(2019){Nelson}, {Springel}, {Pillepich},
  {Rodriguez-Gomez}, {Torrey}, {Genel}, {Vogelsberger}, {Pakmor}, {Marinacci},
  {Weinberger}, {Kelley}, {Lovell}, {Diemer}, \& {Hernquist}}]{Nelson2019}
{Nelson}, D., {Springel}, V., {Pillepich}, A., {et~al.} 2019, Computational
  Astrophysics and Cosmology, 6, 2

\bibitem[{{Pacaud} {et~al.}(2007){Pacaud}, {Pierre}, {Adami}, {Altieri},
  {Andreon}, {Chiappetti}, {Detal}, {Duc}, {Galaz}, {Gueguen}, {Le F{\`e}vre},
  {Hertling}, {Libbrecht}, {Melin}, {Ponman}, {Quintana}, {Refregier},
  {Sprimont}, {Surdej}, {Valtchanov}, {Willis}, {Alloin}, {Birkinshaw},
  {Bremer}, {Garcet}, {Jean}, {Jones}, {Le F{\`e}vre}, {Maccagni}, {Mazure},
  {Proust}, {R{\"o}ttgering}, \& {Trinchieri}}]{Pacaud2007}
{Pacaud}, F., {Pierre}, M., {Adami}, C., {et~al.} 2007, \mnras, 382, 1289

\bibitem[{{Planck Collaboration} {et~al.}(2011){Planck Collaboration},
  {Aghanim}, {Arnaud}, {Ashdown}, {Aumont}, {Baccigalupi}, {Balbi}, {Banday},
  {Barreiro}, {Bartelmann}, {Bartlett}, {Battaner}, {Benabed}, {Beno{\^\i}t},
  {Bernard}, {Bersanelli}, {Bhatia}, {Bock}, {Bonaldi}, {Bond}, {Borrill},
  {Bouchet}, {Brown}, {Bucher}, {Burigana}, {Cabella}, {Cantalupo}, {Cardoso},
  {Carvalho}, {Catalano}, {Cay{\'o}n}, {Challinor}, {Chamballu}, {Chiang},
  {Chon}, {Christensen}, {Churazov}, {Clements}, {Colafrancesco}, {Colombi},
  {Couchot}, {Coulais}, {Crill}, {Cuttaia}, {da Silva}, {Dahle}, {Danese}, {de
  Bernardis}, {de Gasperis}, {de Rosa}, {de Zotti}, {Delabrouille}, {Delouis},
  {D{\'e}sert}, {Diego}, {Dolag}, {Donzelli}, {Dor{\'e}}, {D{\"o}rl},
  {Douspis}, {Dupac}, {Efstathiou}, {En{\ss}lin}, {Finelli}, {Flores-Cacho},
  {Forni}, {Frailis}, {Franceschi}, {Fromenteau}, {Galeotta}, {Ganga},
  {G{\'e}nova-Santos}, {Giard}, {Giardino}, {Giraud-H{\'e}raud},
  {Gonz{\'a}lez-Nuevo}, {Gonz{\'a}lez-Riestra}, {G{\'o}rski}, {Gratton},
  {Gregorio}, {Gruppuso}, {Harrison}, {Hein{\"a}m{\"a}ki},
  {Henrot-Versill{\'e}}, {Hern{\'a}ndez-Monteagudo}, {Herranz}, {Hildebrandt},
  {Hivon}, {Hobson}, {Holmes}, {Hovest}, {Hoyland}, {Huffenberger}, {Hurier},
  {Jaffe}, {Juvela}, {Keih{\"a}nen}, {Keskitalo}, {Kisner}, {Kneissl}, {Knox},
  {Kurki-Suonio}, {Lagache}, {Lamarre}, {Lasenby}, {Laureijs}, {Lawrence}, {Le
  Jeune}, {Leach}, {Leonardi}, {Liddle}, {Linden-V{\o}rnle},
  {L{\'o}pez-Caniego}, {Lubin}, {Mac{\'\i}as-P{\'e}rez}, {Maffei}, {Maino},
  {Mandolesi}, {Mann}, {Maris}, {Marleau}, {Mart{\'\i}nez-Gonz{\'a}lez},
  {Masi}, {Matarrese}, {Matthai}, {Mazzotta}, {Melchiorri}, {Melin}, {Mendes},
  {Mennella}, {Mitra}, {Miville-Desch{\^e}nes}, {Moneti}, {Montier},
  {Morgante}, {Mortlock}, {Munshi}, {Murphy}, {Naselsky}, {Natoli},
  {Netterfield}, {N{\o}rgaard-Nielsen}, {Noviello}, {Novikov}, {Novikov},
  {Osborne}, {Pajot}, {Pasian}, {Patanchon}, {Perdereau}, {Perotto},
  {Perrotta}, {Piacentini}, {Piat}, {Pierpaoli}, {Piffaretti}, {Plaszczynski},
  {Pointecouteau}, {Polenta}, {Ponthieu}, {Poutanen}, {Pratt}, {Pr{\'e}zeau},
  {Prunet}, {Puget}, {Rebolo}, {Reinecke}, {Renault}, {Ricciardi}, {Riller},
  {Ristorcelli}, {Rocha}, {Rosset}, {Rubi{\~n}o-Mart{\'\i}n}, {Rusholme},
  {Saar}, {Sandri}, {Santos}, {Schaefer}, {Scott}, {Seiffert}, {Smoot},
  {Starck}, {Stivoli}, {Stolyarov}, {Sunyaev}, {Sygnet}, {Tauber}, {Terenzi},
  {Toffolatti}, {Tomasi}, {Torre}, {Tristram}, {Tuovinen}, {Valenziano},
  {Vibert}, {Vielva}, {Villa}, {Vittorio}, {Wandelt}, {White}, {Yvon},
  {Zacchei}, \& {Zonca}}]{PlanckIX2011}
{Planck Collaboration}, {Aghanim}, N., {Arnaud}, M., {et~al.} 2011, \aap, 536,
  A9

\bibitem[{{Plummer}(2010)}]{Plummer2010}
{Plummer}, M. 2010, JAGS Version 2.2.0331 user manual

\bibitem[{{Popesso} {et~al.}(2024){Popesso}, {Biviano}, {Bulbul}, {Merloni},
  {Comparat}, {Clerc}, {Igo}, {Liu}, {Driver}, {Salvato}, {Brusa}, {Bahar},
  {Malavasi}, {Ghirardini}, {Robotham}, {Liske}, \& {Grandis}}]{Popesso2024}
{Popesso}, P., {Biviano}, A., {Bulbul}, E., {et~al.} 2024, \mnras, 527, 895

\bibitem[{{Puddu} \& {Andreon}(2022)}]{Puddu2022}
{Puddu}, E. \& {Andreon}, S. 2022, \mnras, 511, 2968

\bibitem[{{Ragagnin} {et~al.}(2022){Ragagnin}, {Andreon}, \&
  {Puddu}}]{Ragagnin2022}
{Ragagnin}, A., {Andreon}, S., \& {Puddu}, E. 2022, \aap, 666, A22

\bibitem[{{Santoni} {et~al.}(2024){Santoni}, {De Petris}, {Ferragamo}, {Yepes},
  \& {Cui}}]{Santoni2024}
{Santoni}, S., {De Petris}, M., {Ferragamo}, A., {Yepes}, G., \& {Cui}, W.
  2024, in European Physical Journal Web of Conferences, Vol. 293, mm Universe
  2023 - Observing the Universe at mm Wavelengths, 00048

\bibitem[{{Serra} {et~al.}(2011){Serra}, {Diaferio}, {Murante}, \&
  {Borgani}}]{Serra2011}
{Serra}, A.~L., {Diaferio}, A., {Murante}, G., \& {Borgani}, S. 2011, \mnras,
  412, 800

\bibitem[{{Stanek} {et~al.}(2006){Stanek}, {Evrard}, {B{\"o}hringer},
  {Schuecker}, \& {Nord}}]{Stanek2006}
{Stanek}, R., {Evrard}, A.~E., {B{\"o}hringer}, H., {Schuecker}, P., \& {Nord},
  B. 2006, \apj, 648, 956

\bibitem[{{Sunyaev} \& {Zeldovich}(1972)}]{Sunyaev1972}
{Sunyaev}, R.~A. \& {Zeldovich}, Y.~B. 1972, Comments on Astrophysics and Space
  Physics, 4, 173

\bibitem[{{Zarattini} {et~al.}(2022){Zarattini}, {Aguerri}, {Calvi}, \&
  {Girardi}}]{Zarattini2022}
{Zarattini}, S., {Aguerri}, J.~A.~L., {Calvi}, R., \& {Girardi}, M. 2022, \aap,
  668, A38

\bibitem[{{Zarattini} {et~al.}(2023){Zarattini}, {Aguerri}, {Tarr{\'\i}o}, \&
  {Corsini}}]{Zarattini2023}
{Zarattini}, S., {Aguerri}, J.~A.~L., {Tarr{\'\i}o}, P., \& {Corsini}, E.~M.
  2023, \aap, 676, A133

\end{thebibliography}

\appendix
\section{Caveats on the filament catalogue: the case of CL1011}
\label{sec:CL1011}

\begin{figure}
    \centering
    \includegraphics[width=0.4\textwidth,trim=20 0 0 -20]{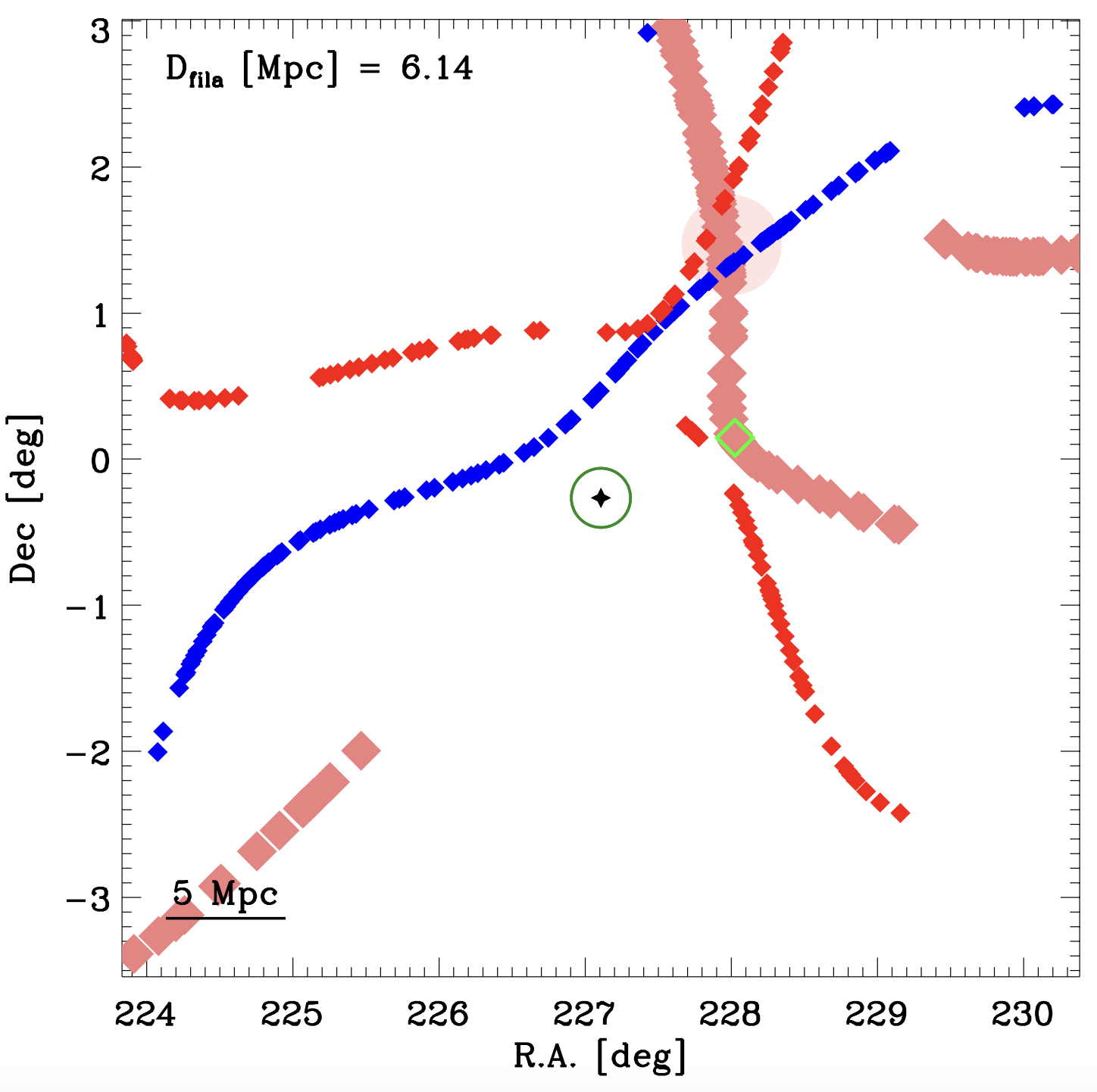}
    \caption{The large scale structure around CL1011. Filaments and intersections in the correct redshift slice are reported in light coral and rose, as in Fig. \ref{fig:example}. The green rhombus is the point to which the distance was measured. We also plot the filaments of the previous and following redshift slices (in blue and red, respectively), in order to look for filaments along the line of sight (e.g. extensions of filaments across different slices). None is found.}
    \label{fig:CL1011}
\end{figure}

As mentioned in Sect. \ref{sec:sample_lss}, the catalogue from \citet{Chen2016} is based on the SDSS DR12 data. These data are homogeneous on a large-scale basis, but can suffer for incompleteness due to various caveats (e.g. fibre collisions, presence of bright stars that exclude areas from the spectroscopic follow up) that can lead to an incomplete detection of filaments. For the cluster CL1011, that is the gas-poor cluster with the largest distance to filaments in our sample, we notice a truncated filament that could become the nearest one, if extended towards the cluster. A possible explanation is that the filament is aligned along the line of sight. In this case, due to the construction of the filament catalogue, we expect the missing part of the filament to be found in one of the contiguous redshift slices. In Fig. \ref{fig:CL1011} we show the large-scale structure of CL1011, including the filaments in the previous and following redshift slices (in blue and red, respectively), to understand the tridimensional structure of this region. 
There are two filaments, one in the former and one in the latter slice, but no one is the extension of the truncated filament.
Of course, the filament could be missing due to incompleteness in the SDSS spectroscopy or in a caveat in the detection code. However, given the lack of concrete evidence, we must rely on the \citet{Chen2016} catalogue, as we have done throughout the rest of our work. 

It is worth noticing, however, that CL1011 is the only cluster in our sample for which we suspect a possible problem in the \citet{Chen2016} catalogue of filament, so we do not think that this issue is affecting the general result. However, it happens to coincide with the system at the largest distance, the only one outside the triangular shape that we find in Fig. \ref{fig:SBx}. In particular, the distance that we measured for the second most-distant cluster ($D_{\rm fila} = 4.6$ Mpc) seems robust.

\section{Modeling of the distance from the filament vs X-ray brightness distribution}
\label{sec:dicotomy}
\begin{figure}
    \centering
    \includegraphics[width=0.48\textwidth,trim=30 -20 60 25]{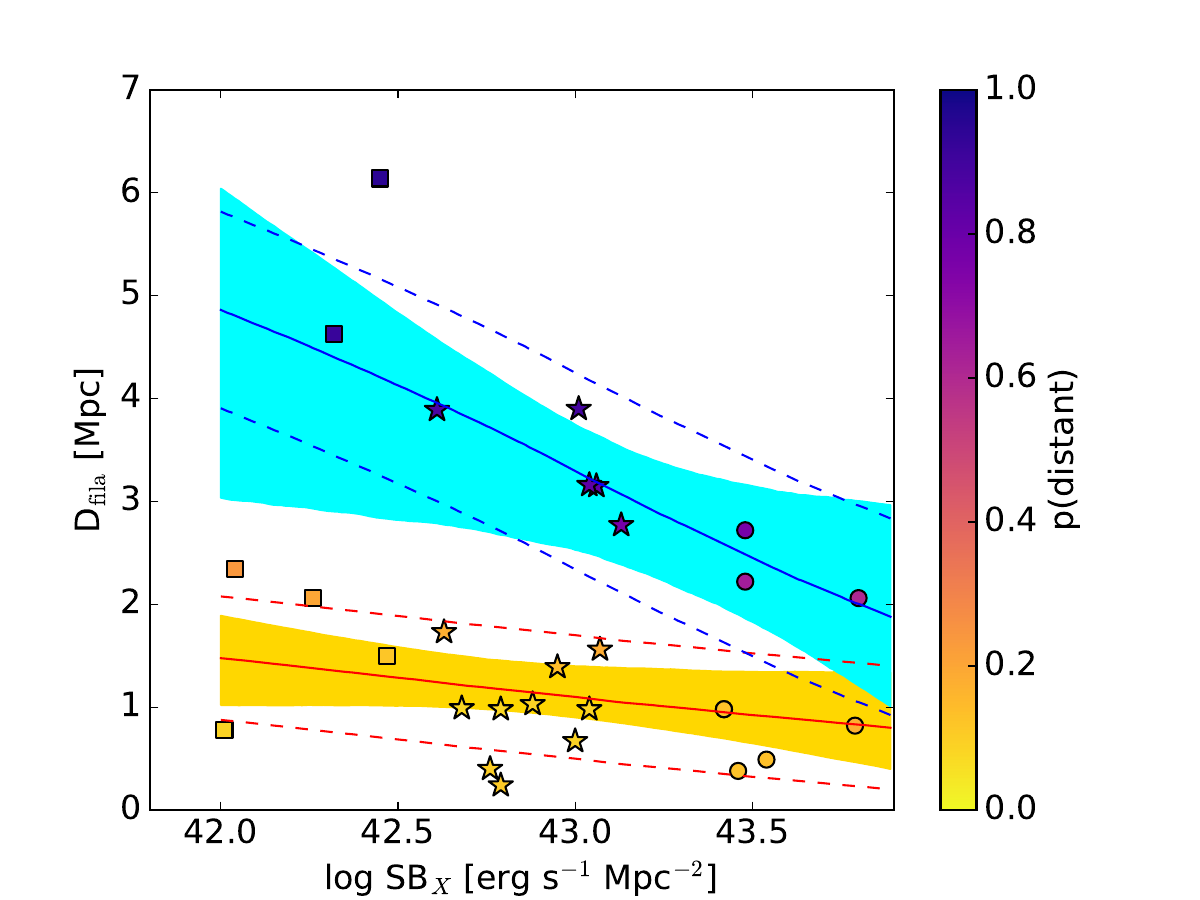}
    \caption{Distance between our clusters and the nearest filament as a function of their X-ray surface brightness SB$_{\rm X}$. 
    Points are colour-coded according to probability to belonging to the more distant population. Shading indicates the 68\% error on the fit, whereas the dashed corridors indicates $\pm 1$ the estimated scatter around the mean relation.}
    \label{fig:SBxv2}
\end{figure}

We notice that there is a possible dichotomy in the distribution of points of Fig. \ref{fig:SBx}. In fact, there is a sort of diagonal gap in that figure.
We therefore fitted the whole data with a mixture of two linear regressions with a Student-t scatter (with 10 degrees of freedom) around them. The latter is taken because it is robust to outliers \citep{Andreon2010}. We took uniform priors for the probability of belonging to the population with lower distance, for the scatters, the intercepts, and the angles \citep[i.e. a Student-t distribution with 1 degree of freedom on the slopes,][]{Andreon2010}. We fit all the data, without assuming which points are drawn from which population, and we sample the posterior with a Gibb sampler \citep[JAGS,][]{Plummer2010}. For parameter identifiability, the second population is the one with larger distance (i.e. with larger intercept, evaluated at 43.0 erg s$^{-1}$ Mpc$^{-2}$). 
Fig.~\ref{fig:SBxv2} illustrates that most points quite neatly split in two nearly separated populations ($p$ is $\sim0$ or $\sim1$, i.e., yellow or blue) and that the two relations have clearly different intercepts. In other terms, the data can be successfully divided in two unmixed populations.
However, we did not find any particular relation characterising the two subsamples in terms of halo mass, the magnitude of the BCG, the magnitude difference between the two brightest members, and the X-ray
temperature. Therefore, we conclude that the observed dichotomy, though successfully modeled, either results from the small sample size or is real but has no impact on the features discussed above. The model in which both slopes are set to zero—meaning that the distance from the filament is independent of the cluster brightness—can only be rejected at approximately 1.5$\sigma$.

\end{document}